\documentclass[conference]{IEEEtran}
\IEEEoverridecommandlockouts
\usepackage{cite}
\usepackage{amsmath,amssymb,amsfonts}
\usepackage{algorithmic}
\usepackage{graphicx}
\usepackage{textcomp}
\usepackage{color}
\def\BibTeX{{\rm B\kern-.05em{\sc i\kern-.025em b}\kern-.08em
    T\kern-.1667em\lower.7ex\hbox{E}\kern-.125emX}}
\begin{document}

%\title{Paper Title*\\

\title{High-resolution antenna near-field imaging and sub-THz measurements with a small atomic vapor-cell sensing element}

\author{\IEEEauthorblockN{David A. Anderson\IEEEauthorrefmark{1}\IEEEauthorrefmark{2}, Eric Paradis\IEEEauthorrefmark{2}\IEEEauthorrefmark{4}, Georg Raithel\IEEEauthorrefmark{2}\IEEEauthorrefmark{3}, Rachel E. Sapiro\IEEEauthorrefmark{2}, and Christopher L. Holloway\IEEEauthorrefmark{5}}
\IEEEauthorblockA{Rydberg Technologies, Ann Arbor, MI USA\IEEEauthorrefmark{2}\\University of Michigan, Ann Arbor, MI USA\IEEEauthorrefmark{3}\\Eastern Michigan University, Ypsilanti, MI USA\IEEEauthorrefmark{4}\\National Institute for Standards and Technology, Boulder, CO USA\IEEEauthorrefmark{5}\\
\IEEEauthorrefmark{1}dave@rydbergtechnologies.com}}
\maketitle

\begin{abstract}
Atomic sensing and measurement of millimeter-wave (mmW) and THz electric fields using quantum-optical EIT spectroscopy of Rydberg states in atomic vapors has garnered significant interest in recent years towards the development of atomic electric-field standards and sensor technologies.  Here we describe recent work employing small atomic vapor cell sensing elements for near-field imaging of the radiation pattern of a K$_u$-band horn antenna at 13.49~GHz.  We image fields at a spatial resolution of $\lambda/10$ and measure over a 72 to 240~V/m field range using off-resonance AC-Stark shifts of a Rydberg resonance.  The same atomic sensing element is used to measure sub-THz electric fields at 255~GHz, an increase in mmW-frequency by more than one order of magnitude.  The sub-THz field is measured over a continuous $\pm$100~MHz frequency band using a near-resonant mmW atomic transition.
\end{abstract}


\section{Introduction}
Sensing and measurement of millimeter-wave (mmW) and THz electric fields with atoms provide advantages over traditional antenna and solid-state detector technologies, including small sensor sizes, absolute measurement capability (SI-based measurement), higher accuracy and precision, as well as re-calibration-free operation, affording greater long-term measurement stability and reliability.  Atomic sensors for electric fields are based on quantum-optical spectroscopy of atomic Rydberg states in vapors contained in spectroscopic cells~\cite{Mohapatra.2007}.  The approach has been developed as a practical means to exploit the sensitivity of Rydberg atoms~\cite{Gallagher} to electric fields over a wide frequency range, from the tens of MHz into the sub-THz regime~\cite{Sedlacek.2012,Holloway2.2014,Simons.2016,Anderson.2016,Miller.2016}, and has garnered significant interest at NIST and National Metrology Institutes worldwide for the establishment of new atomic standards for electric fields~\cite{Gordon.2014,Holloway.2014,HollowayEMC.2017}, as well as in industry for the development of quantum electric-field sensing, measurement, and imaging technologies~\cite{Anderson.2017,RydbergTech}.  In the ongoing pursuit of Rydberg-based electric-field sensors and instrumentation for applications, the development and implementation of small atomic vapor-cell sensing elements suitable for high-spatial-resolution detection of mmW and sub-THz electric fields is desired.  In this report we present recent work employing a small rubidium (Rb) vapor cell to image the electric-field radiation pattern of a K$_u$-band horn antenna.  We demonstrate imaging at a $\sim\lambda/10$ spatial resolution for 13.4884~GHz fields over a dynamic range from 72 to 240~V/m.   The Rydberg-based electric-field measurements are obtained by optical detection of field-induced AC-Stark shifts of an atomic Rydberg resonance.  Measured field profiles are in good agreement with calculated distributions.   We also demonstrate the use of the vapor-cell sensing element for measurements of 255~GHz electric fields over a $\pm$100~MHz continuous frequency band using near-resonant Autler-Townes splittings of a Rydberg resonance.  This is among the highest mmW frequencies measured to date using the Rydberg-atom-based approach; the measurement also demonstrates that a single sensing element can serve a frequency range covering $>4$ octaves.  The presented work serves to pave the way towards practical Rydberg-based field-sensing elements, detectors, and probes that are small and capable of frequency coverage to 1~THz.

%Real-time near-field terahertz imaging with atomic optical fluorescence
%Wade, C.G. 1 ; Sibalic´, N. 1 ; de Melo, N.R. 1 ; Kondo, J.M. 1 ; Adams, C.S. 1 ; Weatherill, K.J. 1
%Source: Nature Photonics, v 11, n 1, 40-3, Jan. 2017;  ISSN: 1749-4885;  DOI: 10.1038/nphoton.2016.214; Publisher: Nature %Publishing Group, UK

\section{mmW detection with atomic vapor-cell sensing elements}
In our experiments, the sensing element comprises a glass cell filled with a Rb vapor.  The active sensing region of the cell has a 3-mm inner dimension; an elongated reservoir with a sample of Rb metal is located several centimeters away.  Two counter-propagating lasers at wavelengths of 780~nm and 480~nm are focused and counter-propagated through the active region of the vapor cell for EIT spectroscopy on field-sensitive Rydberg states.  The 780~nm laser is frequency stabilized to the Rb 5S$_{1/2}$ to 5P$_{3/2}$ transition, whose absorption through the atomic medium is measured, as the 480~nm laser frequency is scanned linearly across a range of 5P$_{3/2}$ to Rydberg-state transitions. The selection of states and the energy ranges scanned are adapted to the field-strength and frequency ranges of the mmW fields to be measured.  When the 480~nm laser frequency is tuned into resonance with a transition into a mmW-field-perturbed Rydberg state, the laser mixes the 5P and the Rydberg state. This leads to a destructive quantum interference of excitation pathways that, over a narrow frequency range at the center of the resonance, results in an increased transmission for the 780~nm light through the vapor cell. These transmission peaks serve to locate the energy levels of the mmW-field-perturbed Rydberg levels. The mmW-field-induced energy shifts of the Rydberg levels as well as the measured splitting patterns of the spectroscopic lines provide an excellent measure for the field strength of the mmW that causes the atomic perturbation.

\section{Near-field imaging of a horn antenna}
We employ the sensing element to measure the near-field of a K$_u$-band 0.695 x 1 inch pyramidal horn antenna emitting 13.4884~GHz fields (near-field range = 58~mm).  The horn is positioned at a distance z=7.5~mm from the sensing element and is translated in steps of $1.9\pm0.2$~mm across the long axis of the horn in the xy-plane.

\begin{figure}
\centering
\includegraphics[width=0.5\textwidth]{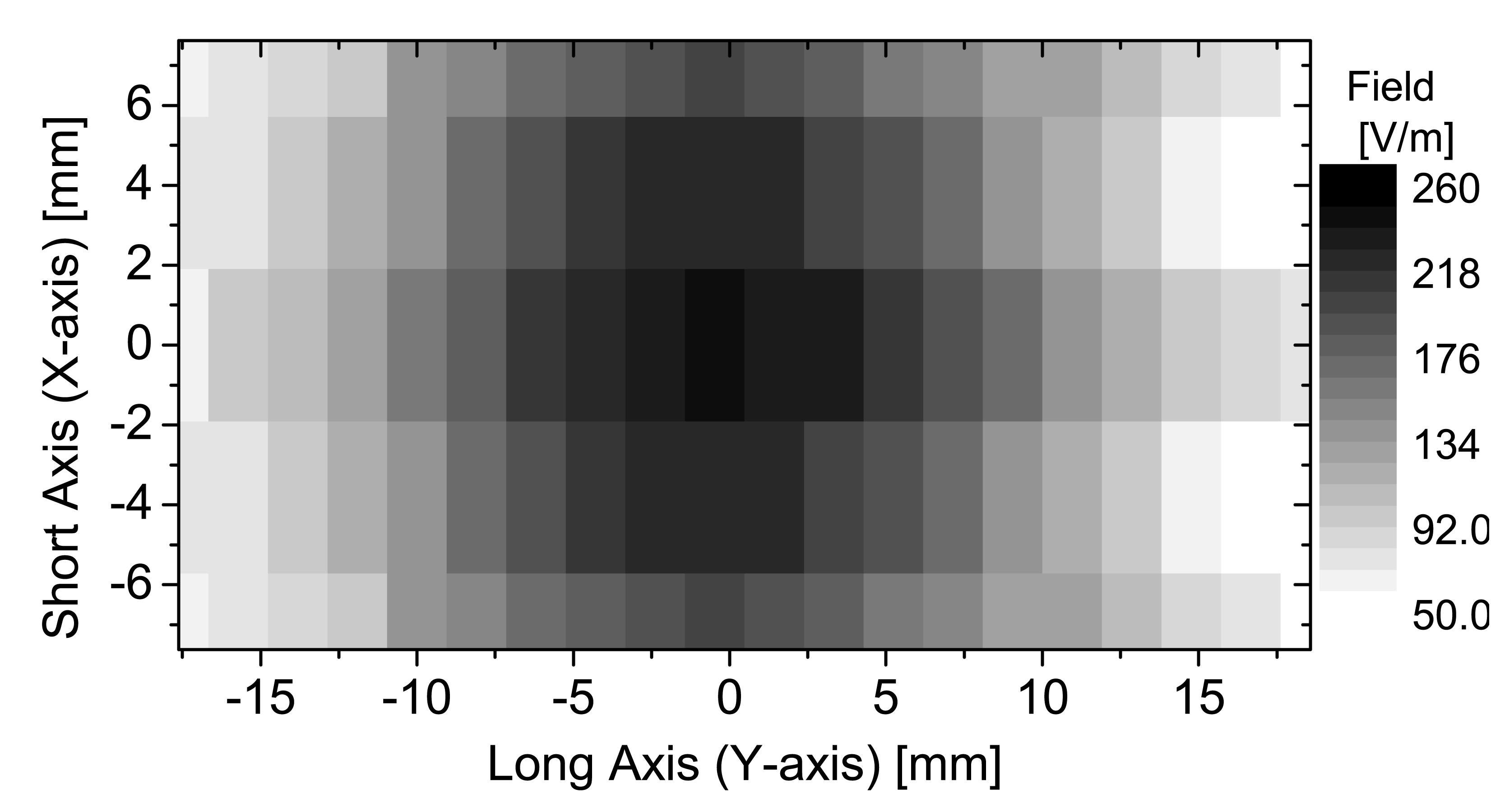}
\caption{Two-dimensional electric field distribution of a horn antenna measured with an atomic sensing element.  The aperture size of the horn antenna is indicated by the white dashed line.}
\label{fig:1}
\end{figure}

\begin{figure}
\centering
\includegraphics[width=0.5\textwidth]{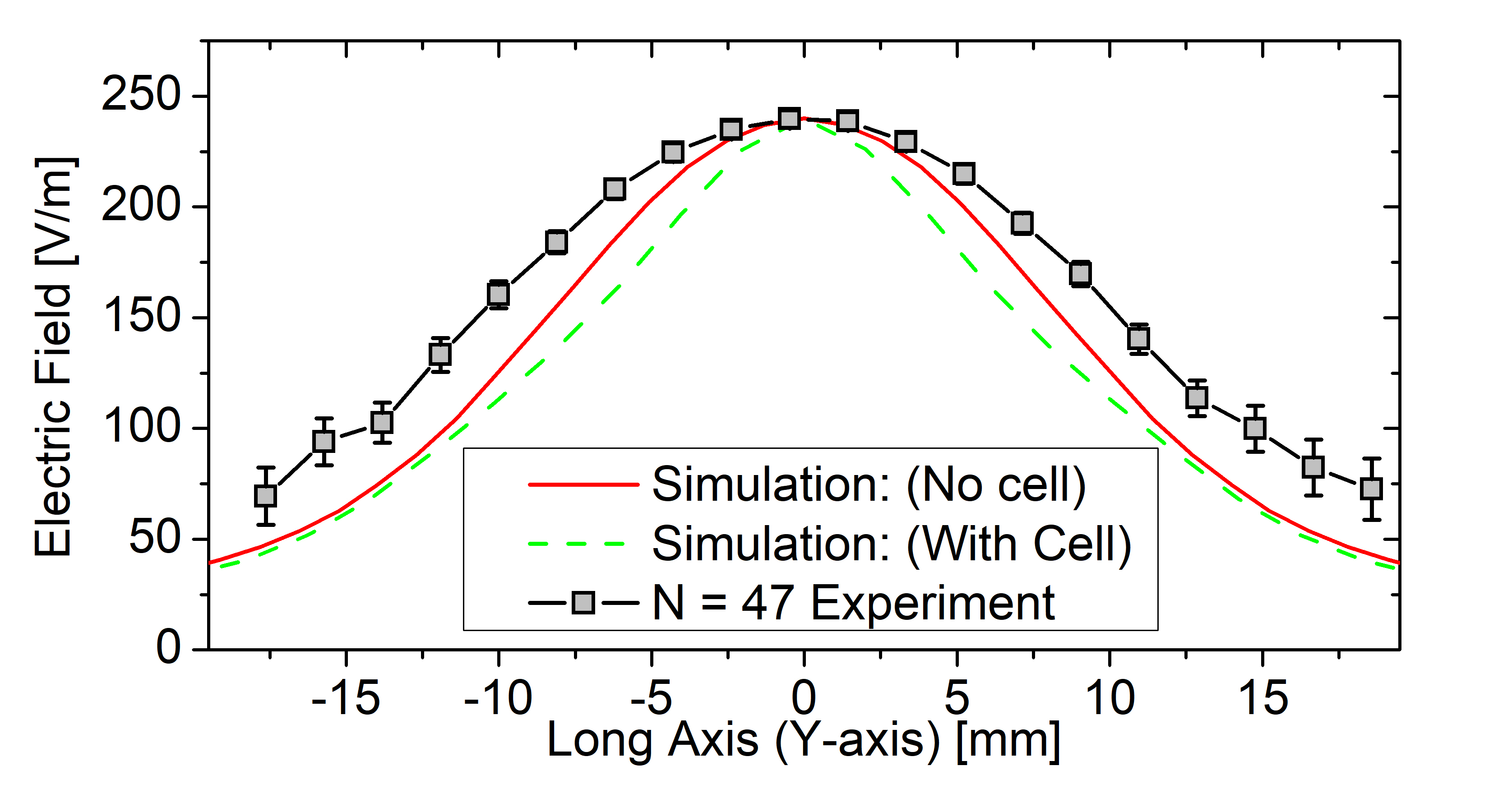}
\caption{One-dimensional electric field distribution along the long-axis of the horn at the short-axis zero for the measurement in Fig.~\ref{fig:1} (gray), calculated distribution in free space (red), and calculated distribution with a dielectric glass cell (dashed green).  The error bars in the measurement correspond to a 1~MHz experimental uncertainty of the atomic line shift.}
\label{fig:2}
\end{figure}

Fig.~\ref{fig:1} shows the two-dimensional spatial electric-field distribution.  The image is composed of a field distribution measured over the top half of the plane and its mirror-image in the bottom half, symmetric across the short-axis zero.  The electric field at each position of the horn is obtained from EIT detection of the mmW-induced AC-Stark shift of the 47S$_{1/2}$ Rydberg state using the relation $E=(4 \Delta/\alpha)^{0.5}$, where $\Delta$ is the measured peak shift of the atomic line, and $\alpha=4.099\times10^{-3}$~MHz/(V/m)$^2$ is the Rb 47S AC polarizability at 13.5~GHz.  This relation holds in fields low enough that the field does not induce transitions, and that higher-order, non-quadratic shifts of the Rydberg level are not significant.  In Fig.~\ref{fig:2} we plot the measured one-dimensional field profile across the long axis at the short-axis zero in Fig.~\ref{fig:1} with calculated field distributions with and without a dielectric cell structure.  We obtain overall good agreement between measured and calculated distributions.  The slight increase in field measured at the wings of the distribution compared to the calculated distributions may be attributed to the imperfect modeling of detailed geometry and the material of the glass cell that holds the atomic vapor, and to the dielectric mount of the cell that is not accounted for in the present calculations.  The detailed characterization of vapor-cell sensing elements on mmW fields for measurements with atomic sensors is a topic of on-going and future work.

\section{Sub-THz electric-field measurements with millimeter vapor cells}
A key advantage of atomic electric-field sensors is that a single small vapor-cell sensing element can be employed in measurements over a wide mmW frequency range. Here we perform measurements of electric fields at 255.2851~GHz using the same atomic vapor-cell sensing element.  This is more than four octaves higher in frequency compared to the previous measurements.  The measurements are performed in the far-field of a conical mmW horn emitting 255.2851~GHz sub-THz radiation field that is resonant with the 27S$_{1/2}$ to 26P$_{3/2}$ electric-dipole transition in Rb. The mmW frequency is varied around the atomic resonance for field measurements over a 200-MHz wide band.

\begin{figure}
\centering
\includegraphics[width=0.5\textwidth]{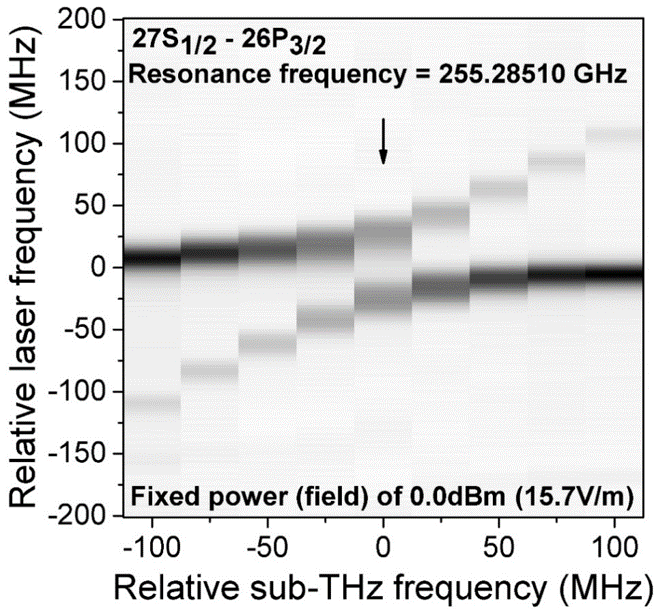}
\caption{Atom-based measurements of electric fields over a frequency range 255.185~GHz to 255.385~GHz.}
\label{fig:3}
\end{figure}

Fig.~\ref{fig:3} shows an experimental atomic spectral map where the 780nm laser transmission through the sensing element (gray-scale; dark = increased transmission) is plotted as a function of the relative 480~nm laser frequency (Y-axis) and the relative mmW frequency (X-axis). There, the laser frequency is relative to the mmW-field-free 27S$_{1/2}$ Rydberg resonance, and the mmW frequency  relative to the 27S$_{1/2}$ to 26P$_{3/2}$ transition frequency.
The mmW power injected into the mmW transmission line is fixed at 0~dBm.  The resonance condition is indicated by the arrow in the plot, at the center of the avoided crossing between two dressed atom-field states. As the sub-THz frequency is tuned away from the atomic resonance, the two Autler-Townes-split spectral lines, whose on-resonance splitting is directly proportional to the mmW electric field, separate further, with a splitting given by the effective Rabi frequency $\Delta f$.  As the detuning of the sub-THz radiation is increased, the splitting of these off-resonant Autler-Townes lines is then given by $\Delta f^2 = (\Delta f_0)^2 + \delta^2$,
where $\Delta f_0$ is the resonant Autler-Townes frequency splitting and $\delta$ is the detuning of the sub-THz field from the atomic resonance.  In Fig.~\ref{fig:3}, the resonant splitting corresponds to a mmW field of 15.7~V/m, and $\delta$ is varied from -100~MHz to +100~MHz about the resonance frequency. The frequency range over which the measurement method is effective in measuring the field has a lower limit given by the range over which both Autler-Townes lines are visible, and an upper limit given by the range within which the dominant Autler-Townes component exhibits a detectable AC Stark shift. In the case of Fig.~\ref{fig:3} the frequency range over which the measurement method is effective is at least 200~MHz.

\section{Conclusion}
We report on near-field imaging of the radiation pattern of a horn antenna at 13.49~GHz using a small atomic vapor cell sensing element with a resolution of $\sim\lambda/10$ and covering a dynamic field range from 72 to 240~V/m using off-resonant AC-Stark shifts of a Rydberg resonance.  The imaging results are in good agreement with calculations.  The same atomic sensing element is used for sub-THz electric field measurements at 255~GHz, more than one order of magnitude higher in mmW-frequency.  The sub-THz measurements are performed over a 200~MHz-wide frequency band centered on a resonant atomic transition.

\section*{Acknowledgment}
This work is supported by Rydberg Technologies, and by the Defense Advanced Research Projects Agency (DARPA) and the Army Contracting Command-Aberdeen Proving Ground (ACC-APG) under Contract No. W911NF-15-P-0032.

%\bibliographystyle{IEEEtran}
%\bibliography{IEEEabrv,GSMM_bibfile}

\end{document}